\documentclass[doublespacing]{elsart}

\usepackage{natbib}

 \usepackage{graphicx}
\usepackage{epsfig}
\journal{Astroparticle Physics}
\usepackage{amssymb}

\begin{document}

\begin{frontmatter}















\title{Determination of the Night Sky Background around the Crab pulsar using
  its optical pulsation}





\author[Potch,Complu]{E. O\~na-Wilhelmi}
\ead{emma@gae.ucm.es}
 \author[IFAE]{J. Cortina}
\ead{cortina@ifae.es}
\author[Potch]{O.C. de Jager}
\ead{fskocdj@puk.ac.za}
\author[Complu]{V. Fonseca} 
\ead{fonseca@gae.ucm.es}

\address[Potch]{Unit for Space Physics, Northwest University, Potchefstroom 2520, South Africa}
\address[Complu]{Dept. de F\'{i}sica At\'omica, Molecular y Nuclear, UCM,
  Ciudad Universitaria s/n, Madrid, Spain}
\address[IFAE]{Institut de Fisica d'Altes Energies, UAB, Barcelona, Spain}

\begin{abstract}
The poor angular resolution of imaging $\gamma$-ray telescopes is offset by
the large reflector areas of next generation telescopes such as MAGIC (17~m
diameter), which makes the study of optical emission associated with some
$\gamma$-ray sources feasible.
Furthermore, the extremely fast time response of photomultipliers (PMs) makes them
ideal detectors for fast (subsecond) optical transients and periodic sources
like pulsars. The optical pulse of the Crab pulsar was detected with the HEGRA
CT1 central pixel using a modified PM, similar to the future MAGIC camera PMs. 

The HEGRA CT1 telescope is a 2.1~m radius \^{C}erenkov telescope,
which has seen the Crab optical pulsations. The corresponding observation time
required for a detection is 200 seconds, but MAGIC will only require $\sim
30$~sec if the same strategy 
with standard PMs is used. 

The constancy of the pulsed fraction (with a relatively small error) during
the 2 hours CT1 measurements shows that the pointing did not change during the
observations. The purpose of these periodic observations was to
determine the light of the night sky (LONS) for the galactic anticenter Crab region, and
to compare it with the LONS for a nearby bright star ($\zeta$ Tauri). These
obtained LONS values vary between 4.6 and 4.3 $\times 10^{12}\; ph \cdot
m^{-2} \cdot s^{-1} \cdot sr^{-1}$ (with a coarse estimated error of $10\%$), being the first value from the timing
signature of the Crab pulsar, and the second one was derived from the
technique using the $\zeta$ Tauri standard star. Our results are between 2.5
and 3 times larger than the previously measured LONS (outside the galactic
plane) of $(1.7\pm 0.4)\times 10^{12} ph\cdot m^{-2} \cdot s^{-1} \cdot
sr^{-1}$, as expected since the Crab pulsar is in the galactic plane, which
implies a slightly higher energy threshold for Crab observations, if the
higher value of CT1 measured LONS rate for galactic sources is used.

\end{abstract}


\end{frontmatter}


\section{Introduction}

Imaging Atmospheric \^{C}erenkov Telescopes (IACTs) can be used to detect the
optical emission of an astronomical object through the increased DC current of
the camera pixels. Each of these pixels has a field-of-view (FOV) in the range
of 0.1-0.3$^{\circ}$. Hence, when an IACT follows a star, its optical emission
is fully contained in the telescope central pixel.

The Crab Nebula is known to be the most important calibration source for high
energy astrophysics, and especially for the ground-based atmospheric
IACTs. The night sky background (LONS) influences the trigger threshold, the
energy and the spectral determination of $\gamma$-rays, therefore the LONS
around the Crab Nebula must be accurately known in the wavelength interval of
the photomultipliers (PMs) used for the IACT cameras. Mirzoyan and Lorenz
\cite{Razmik} measured the LONS at the site of the HEGRA experiment for a
narrow angle, using a detector attached to the telescope CT2, which integrates
only the starlight from the background of faint stars, obtaining a value of
$I_{LONS}$=$(1.7\pm 0.4)\times~10^{12}~ph\cdot m^{-2}\cdot s^{-1}\cdot sr^{-1}$ (outside the galactic plane). In this paper we measure this
$I_{LONS}$ for the Crab region and we compare it with the value of
\cite{Razmik}, using a modified PM in the central pixel of the CT1 camera.

The optical pulsed emission of the Crab pulsar has
already been detected using IACTs \cite{whipple_crab} and 
other \^{C}erenkov detectors \cite{celeste_crab}. The main purpose
of such studies are to determine the timing parameters of the
pulsar, with which the $\gamma$-ray arrival time can be folded, thus
reducing the search for pulsed emission to a single statistical trial.
It also allows us to test the accuracy of the used timing system and
the solar system barycentric correction software, when comparing the timing
parameters against contemporary radio data.
This paper reports the optical timing signature of the Crab pulsar obtained, using the
central pixel of CT1 \cite{CT1}, and the DC observations of the bright
nearby star $\zeta$ Tauri, used to derive the LONS around the Crab
Nebula. These two values cross-check the LONS measurements obtained using two
different techniques: temporal (Crab pulsation) versus spatial (nearby star).

Our results with CT1 can be extended to a much larger telescope such as MAGIC
\cite{MAGIC} (17 meter diameter). IACTs are calibrated using as reference the
Crab signal, 
and we can therefore use this regular Crab monitoring, 
not only for \^{C}erenkov measurements, but also to measure the optical Crab pulsation
with a modified PM in the telescope camera. Furthermore, since the
astronomical community does not monitor the Crab pulsar continuously (on a 24
hour timescale) in radio or optical, observations with \^{C}erenkov telescopes
will also help to serendipitously discover pulsar glitches or period irregularities.

\section{The Crab Optical Spectrum and Expected Single Photoelectron Response.}

The optical spectra of the Crab pulsar main pulse and interpulse, and the
spectrum of the underlying nebula \cite{carram}, follow the power laws given
in table 1. The nebula with a size of $4'\times 6'$ fits well into a single
pixel.

\begin{table}[ht]
\footnotesize
\begin{center}
\begin{tabular}{|l|c c|}
\hline
Flux & 
\parbox{1.5cm}{\centering\scriptsize K} &
\parbox{1.5cm}{\centering\scriptsize $\alpha$} \\ 
\hline

{\scriptsize Main pulse} & $5.9 \times 10^{-15}$   & +0.2 $\pm$ 0.1 \\

{\scriptsize Interpulse} & $1.9 \times 10^{-15}$ & +0.2 $\pm$ 0.1 \\

{\scriptsize Underlying nebula} & $8 \times 10^{-12}$ & -0.4 $\pm$ 0.1 \\

\hline

\end{tabular}
\end{center}
\caption[\it Note]
{\it Flux in the 0.1 phase window centered on the main pulse and
  interpulse, where,
  $F_{\lambda}$=$K\times(\lambda$/$\lambda_{o})^{-\alpha-2}$, $\lambda_{o}$ =
  6000 $\AA$, 5000 $\AA$ $<$ $\lambda$ $<$ 7500 $\AA$. The useful wavelength
  range is 3000-8000 $\AA$. Fluxes are in $erg~\cdot~s^{-1}\cdot\AA^{-1}\cdot~cm^{-2}$.}
\label{Fig1-1}
\end{table}



We can calculate the single photoelectron rate for CT1 and MAGIC. Note that
the poor angular resolution (a single pixel with FOV of 0.25$^{\circ}$ for CT1
and 0.1$^{\circ}$ for MAGIC), is compensated for by the large mirror area A
(10~$m^{2}$ for CT1, 250~$m^{2}$ for MAGIC). 

The expected photoelectron rates are calculated with the following expression:
\begin{equation}
R = {\int F_{\lambda} \cdot \eta \cdot Q_{\nu} \cdot A~\cdot \beta~d\nu},
\label{eq1}
\end{equation}
where $\eta$ is the mirror reflectivity and $Q_{\nu}$ the 
frequency dependent quantum efficiency of the PMs. The parameter $\beta$
takes into account the photoelectron detection on the first dynode (0.9) and
the light guide efficiency (90$\%$).

For the Crab pulsar, the theoretically expected rates (Eq. \ref{eq1}) are
given in Table \ref{table 2}, for CT1 and MAGIC telescopes.

\begin{table}[ht]
\footnotesize
\begin{center}
\begin{tabular}{|c c c c c|}
\hline
Telescope & 
\parbox{2.5cm}{\centering\scriptsize Total Pulsar Rate \\ (phe $\cdot s^{-1}$)} &
\parbox{2.5cm}{\centering\scriptsize Nebula Rate \\ (phe $\cdot s^{-1}$)} & 
\parbox{2.5cm}{\centering\scriptsize LONS Rate \\ (phe $\cdot s^{-1}$)} & 
\parbox{2.5cm}{\centering\scriptsize Background Rate \\ (phe $\cdot s^{-1}$)} \\ 
\hline

{\scriptsize CT1} & $1.5 \times 10^{5}$  & $1.4 \times 10^{8}$  & $3.3 \times 10^{7}$ & $1.7 \times 10^{8}$
\\
\hline

{\scriptsize MAGIC} & $3.7 \times 10^{6}$  & $3.5 \times 10^{9}$  & $1.3 \times 10^{8}$ & $3.6 \times 10^{9}$
\\

\hline
\end{tabular} 
\end{center}
\caption[\it Note]
{\it Expected currents (in photoelectrons/s) of the optical 
emission of the Crab pulsar in the central pixel of the 
CT1 and MAGIC telescopes. Background rate means the total background: LONS + Nebula}
\label{table 2}
\end{table}

If we assume that the PM has no inherent fluctuations
(for example the fluctuations of the electron multiplication
process between dynodes),
we may derive the basic scaling parameter $X=p \sqrt(N)$
based on photon (Poissonian) statistics,
for all test statistics, which measures
the integrated difference between the measured pulse profile and a uniform distribution (no pulsed signal)
\cite{jager}. The parameter $p$ is the number of pulsed events, divided by
the total number of events $N$ (signal plus noise). This $X$ also corresponds
approximately to the expected DC excess (measured in Gaussian standard
deviations). Assuming that the ephemeris for folding the arrival times of optical photons
from the Crab pulsar to phases is correct, the expected scaling parameters for
CT1 and MAGIC (taking $I_{LONS}$ from \cite{Razmik}, although we expect
it to be higher), based on photon statistics alone, are: 

\begin{equation}
X_{ct1} = 11 \sqrt{\frac{T}{1~sec}}
\end{equation}

\begin{equation}
X_{MAGIC} = 62 \sqrt{\frac{T}{1~sec}}
\end{equation}

From the basic scaling parameter we infer the expected time to detect the
optical pulsation for a given significance. 
The theoretical rate from the Crab pulsar shows that it would be 
possible to detect the Crab optical pulse with the CT1 telescope 
in T=75~ms, with a significance of 3$\sigma$. For MAGIC, the situation is even
better because the main pulse could be detected within~5 ms, which is the
approximate duration of this pulse. This implies that MAGIC has the
capability to monitor single pulses from Crab, based on
theoretical arguments, and using the LONS intensity corresponding to an
extragalactic source.

This result is only valid if the photodetector 
itself has no noise, and the only source of fluctuations is
Poissonian in nature, due to the signal. The expected detection times increase
dramatically when we take fluctuations into account, as shown by the
measurements presented below.


\section{Measurements}

Measurements were done with the HEGRA CT1 telescope, during February 11-16th
2002 and during November 1-8th 2002. The standalone HEGRA CT1 telescope has a
reflector area of 10~$m^2$ and a camera consisting of 127 PMs, each with
0.25$^{\circ}$ FOV. 
\\
Datasets were obtained with the Crab in the central pixel, and OFF
source runs were also taken to check for any systematics.
For optical observations the central pixel PM was modified. The DC branch,
designed to monitor the DC current of the pixel, was adjusted to detect pulses of $\sim$3~ms (timescale of the pulse width), and the AC branch,
which is designed to transmit the ns fast signals
generated by the \^{C}erenkov showers was removed \cite{tdas02}.

The DC current was measured using a National Instruments
PCI NI 6043E digitising card. The sampling rate was set to 2~kHz and the 
pixel line was AC-coupled to reject low frequencies, and increase 
the card's effective dynamic range.


\section{Analysis and Results of the Pulsed Optical Signal Measurements}

The arrival times $t_i$ were folded to obtain the phases $\Phi_{i}$. 
\begin{equation}
\Phi_{i} = \Phi_{0} + \nu (t_{i} - t_{0}) + \frac{1}{2} \dot{\nu} (t_{i} - t_{0})^2,
\end{equation} 
where $\nu, \dot{\nu}$ are the frequency and frequency derivative, and $t_0$
is the reference epoch.
 
To transform these arrival times to the solar barycentre system, we used the
TEMPO software \cite{tempo}. A correction due to a small average drift in the
digitisation card clock, as measured with respect to an atomic reference clock, was 
also introduced. The Jodrell Bank ephemeris (frequency, first frequency
derivative and reference time) applicable to these observations are given in
Table \ref{table 3} (Jodrell Bank Crab pulsar monthly ephemeris).

\begin{table}[ht]
\footnotesize
\begin{center}
\begin{tabular}{|l|c c c c|}
\hline
\parbox{2.5cm}{\centering\scriptsize Date } &
\parbox{2.5cm}{\centering\scriptsize $t_{o}$ \\ (MJD) } & 
\parbox{2.5cm}{\centering\scriptsize $\nu$   \\ (Hz) } & 
\parbox{2.5cm}{\centering\scriptsize $\dot{\nu}$ \\ ($10^{-15}$ $sec^{-2}$)} &
\\
\hline
{\scriptsize 15th October 2002} & 52562  & 29.8132421622  & -373777.06 &
\\
\hline
{\scriptsize 15th November 2002 } & 52593 & 29.8122410972 & -373758.90  &
\\

\hline
\end{tabular} 
\end{center}
\caption[\it Note]
{\it Radio Timing Parameters for the Crab pulsar for the indicated epochs.}
\label{table 3}
\end{table}
A phaseogram was produced for each independent frequency confined to the wide
frequency range between 26.0 and 32.0~Hz. The independent Fourier spacing
(IFS) between this range of independent frequencies is known to be $1/T$
(periodogram bin size), where T hours is the observation time (2~hours).

\begin{figure}[h]

\epsfysize=7cm 
\epsfxsize=14cm 
\epsfbox{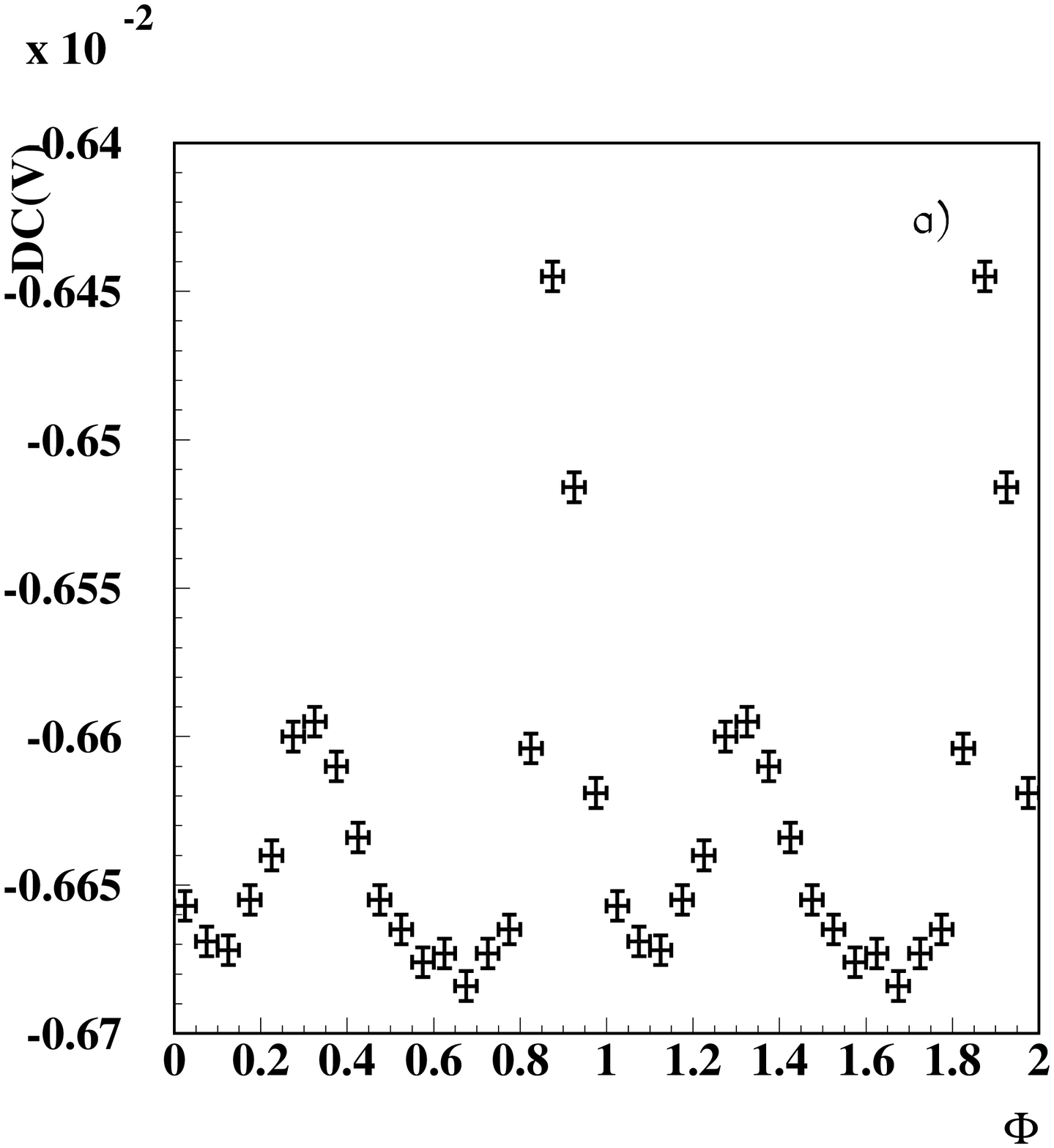} \\
\epsfysize=7cm 
\epsfxsize=14cm 
\epsfbox{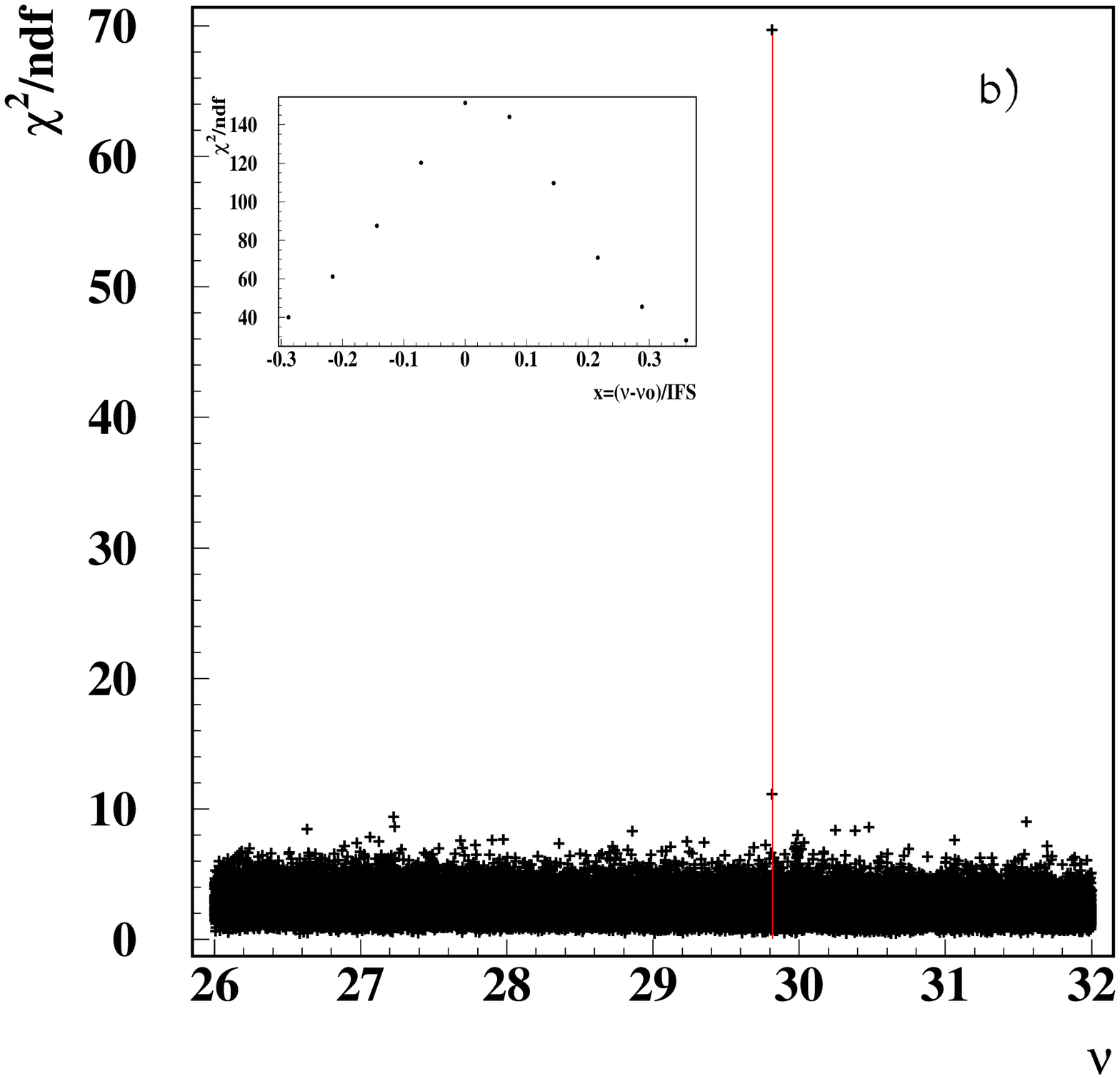}
\caption[h]{\it \small Observations of the optical Crab pulsar with CT1. On
  the top, the two peaks are clearly visible in the light curve with a
  separation in phase of $\sim$ 0.4. On the bottom (b) the excess in the
  reduced $\chi^2$ searching in a wide range of frequencies, using the 15th
  October ephemeris. The difference between the frequency found with 15th October
  ephemeris and 15th November is consistent with the drift in the
  digitization card clock. The inset shows the value of the reduced $\chi^2$
  vs. x = ($\nu - \nu_{o})/IFS$, where $\nu_{o}$ corresponds to 15th October.}

\label{l_curve}

\end{figure}

The folded intensities for each test frequency were tested against a uniform
distribution by performing a $\chi^2$ fit to a constant intensity. The reduced
$\chi^2$ was calculated (Figure \ref{l_curve}). A maximum value of the reduced
$\chi^2$ of $\sim$ 154 was found and the periodogram signal is contained within
one IFS of the expected Jodrell Bank ephemeris, consistent with the
uncertainty in the drift expected from the clock on the digitisation
card. Absolute phase matching with the expected
radio/optical/X-ray/$\gamma$-ray main pulse position was not possible, since
an absolute UTC match was not done, and the drift was not accurately measured
during observations.

Furthermore, the frequency resolution for the 2 hour observation was too
coarse to resolve the differences between the ephemerides corresponding to 15
October and 15 November. The predicted frequency shown in Figure \ref{l_curve}
is therefore consistent with both ephemerides.

Figure \ref{l_curve} shows our measured Crab pulse profile, with two sharp
peaks, with a $\sim 0.42$ phase difference. 
The corresponding pulsed fraction can now be used to isolate the LONS and
cross-check the previous measurement \cite{Razmik}. Using our results we can
infer the MAGIC response for the Crab pulsar.

\section{Determination of the galactic anticenter LONS}

In this section, we will determine the galactic night sky background around
the Crab pulsar by using the pulsed fraction of the signal detected in
CT1. This result will be checked by comparing it with the LONS
calculated from the intensity of the CT1 camera response to $\zeta$ Tauri.
 
It is possible to derive the pulsed fraction of the optical Crab signal from
the density function of the light curve. This density function results from
the unpulsed background (nebula and night sky background) with fraction $1-p$
and the density function $f_{s}$($\theta$) of the pulsed signal coming from
the Crab pulsar:

\begin{equation}
f(\theta) = p f_s(\theta) + (1-p),
\end{equation}
where p is the pulsed fraction, if the density function is normalised to unity. 
The off-pulse interval (region of minimum intensity) following the interpulse
and preceding the main pulse was used to estimate the pulsed fraction.

Taking the proper errors in each bin into account (exploiting the Central
Limit Theorem), we derive a pulsed fraction for the Crab of
$p=(5.00\pm~0.02)\times10^{-4}$.

\begin{figure}[h]

\epsfysize=12cm 
\epsfxsize=12cm 
\hspace{0.5cm}
\epsfbox{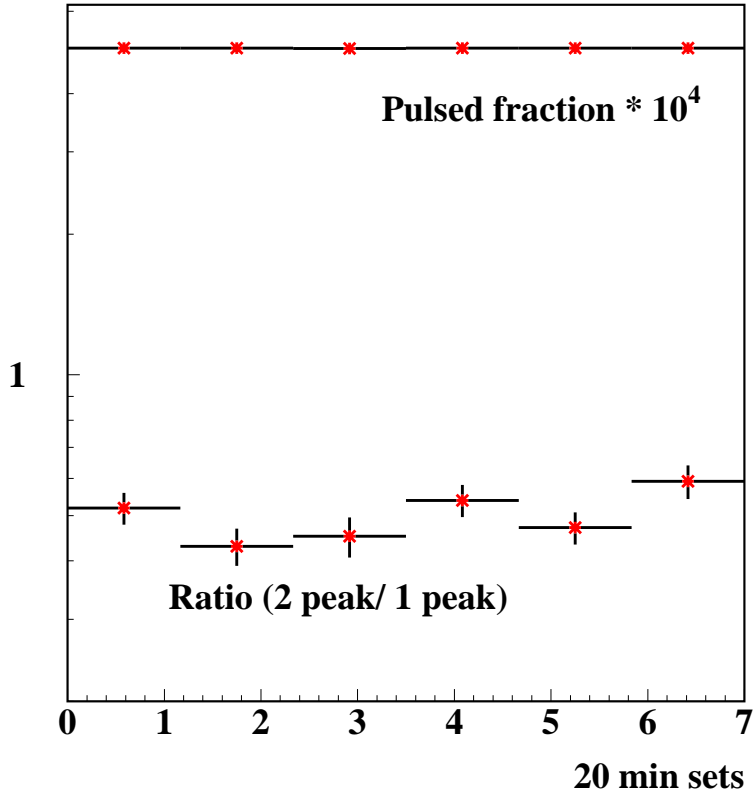}
\caption[h]{\it Test of the telescope response to the pulsed signal in two
  hours. The pulsed fraction times $10^4$ for
  the set of 20 min observation slices is at the top of the plot. The ratio between the first and second peak remains
  constant with time, as shown at the bottom.\ }
\label{time}
\end{figure}

To check the stability of the signal with time, the total data set of two hours
was divided into six data sets of 20~minutes each. The pulsed fraction,
as well as the ratio between the first and second peaks was calculated
for each subset.
The results are shown in the Figure~\ref{time}. It is clear that both
parameters appear to be stable within statistical errors. An indirect
indication of the stability of telescope tracking during two hours
observations is shown by the constant value of the pulsed fraction.

There is a difference between the measured CT1 Crab pulsed fraction of
$5\times10^{-4}$ and the theoretical value given in the previous section ($p_{theor}~=~8.6~\times~10^{-4}$). The discrepancy can be attributed to the
uncertainty on the measurement of the LONS, the influence of the mirror point
spread function (PSF), and the fact that the Crab pulsar was off-center due to
uncorrected but constant misspointing in the telescope.

To determine the galactic LONS around the Crab pulsar, we first need
to know the PSF of the telescope. With this purpose, we used the bright source
$\zeta$ Tauri to measure the amount of light, which falls within the central
pixel. The same source is used to measure the telescope misspointing.

\begin{figure}[h]
 
\epsfysize=8cm 
\epsfxsize=7.5cm 
\hspace{0.5cm}
\epsfbox{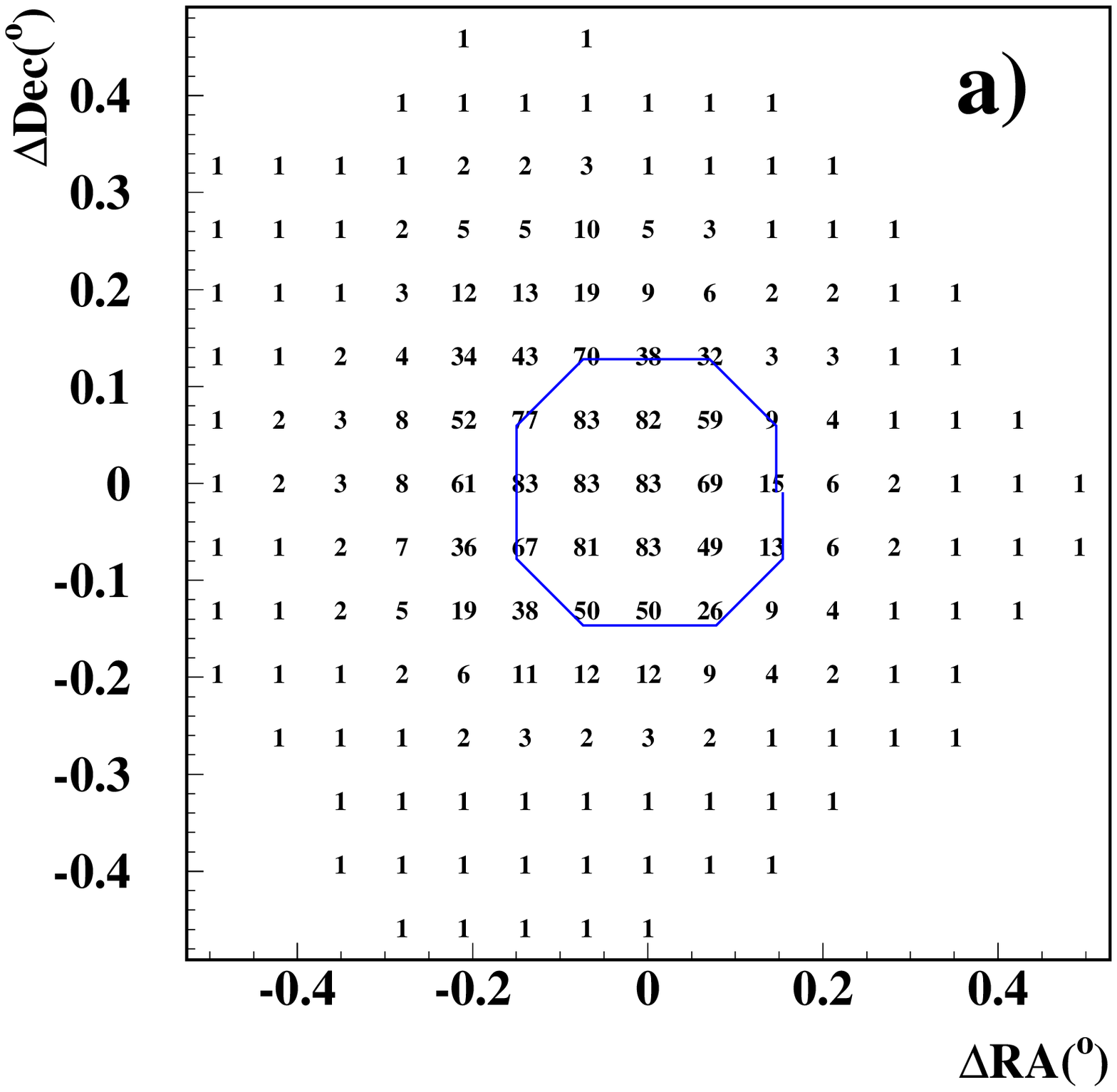}
\epsfysize=9cm 
\epsfxsize=7.5cm 
\epsfbox{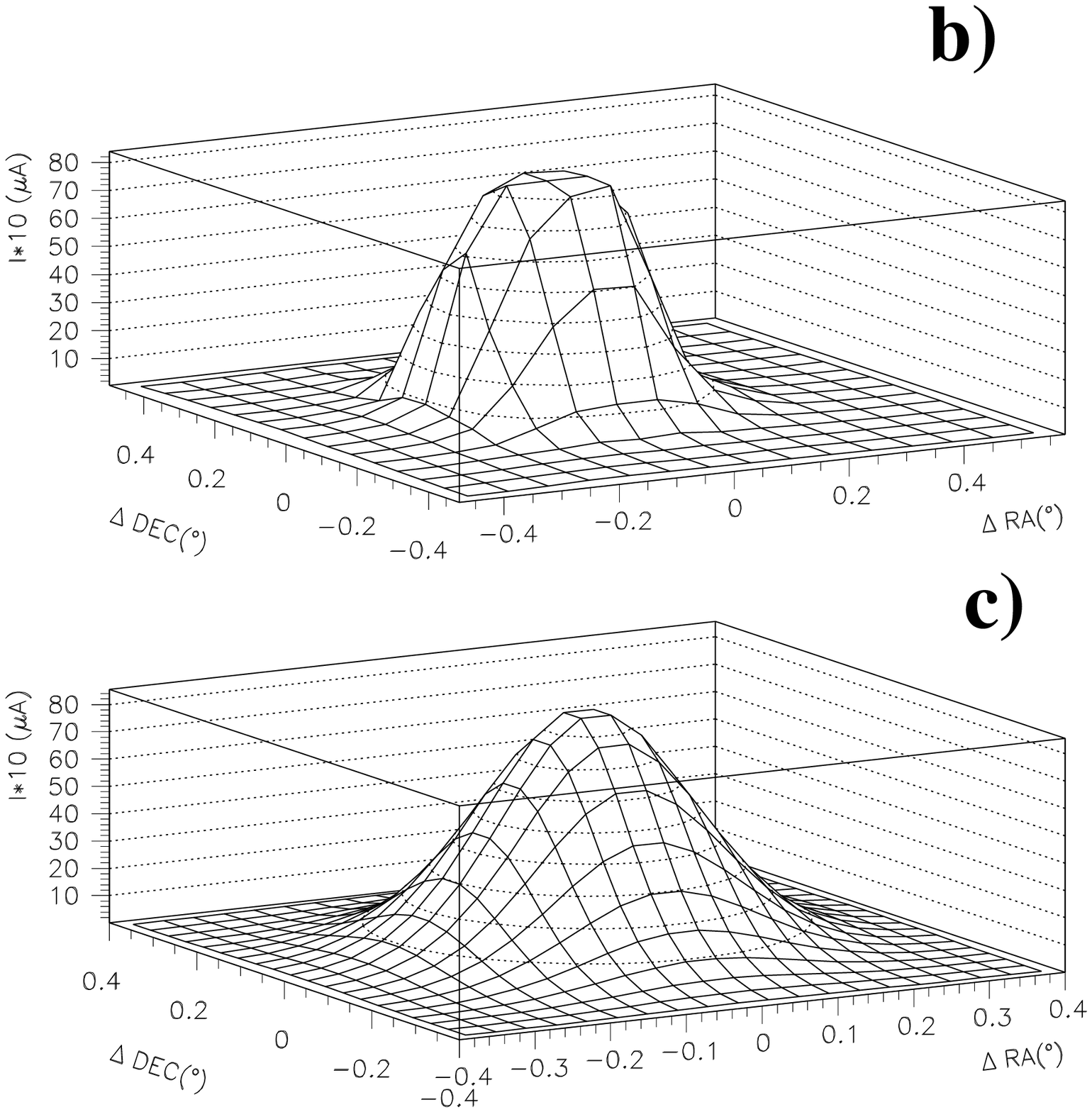}



\caption[h]{\it \small DC currents due to $\zeta$ Tauri on the CT1 central
pixel, in 0.1 $\mu$A units. Left (a): $\zeta$ Tauri signal in the CT1 camera,
with background already subtracted. The zero position corresponds to the true
position of the star in equatorial coordinates (maximum intensity for zero mispointing). The approximate shape of the
central pixel is drawn to illustrate the size of the PSF relative to the pixel
size. Top right: (b) $\zeta$ Tauri observed intensity in a 2-dimensional plot. 
Bottom right: (c) The same signal, fitted with a bidimensional Gaussian.}
\label{zetatauri}
 
\end{figure}

The selected source was $\zeta$ Tauri (ICRS 2000.0 coordinates are 5$^h$:37$'$:38.70$''$ RA,
21$^{\circ}$:08$'$:33.0$''$ DEC with a magnitude of 2.84), which also appears
in the camera field of view when CT1 points to the Crab pulsar. The 
advantages of using this reference star are the following: 
(1) we know its photometric properties and (2) the atmospheric extinction
will be the same for both the pulsar and the star, given their
proximity in the sky.
The PSF of the telescope was determined from a raster scan with a resolution
finer than the 0.25$^{o}$ pixel size (see Fig. \ref{zetatauri}(a)), with zero
position corresponding to the true position of the star). Since the
telescope is focused to $\sim$6~km (maximum \^{C}erenkov shower), the image of
a point-like source will be smeared out around 0.14$^o$.

The signal from $\zeta$ Tauri is slightly shifted to lower right ascension
values, due to the telescope mispointing. This was 
accounted for by calculating 
the maximum value of the
signal, and measuring the difference between the shifted measurement and the true position.  A
schematic figure of the CT1 central pixel is also drawn in Figure 3. To calculate the
true signal collected by the central pixel, the signal intensity was fitted to a bidimensional Gaussian (fig. \ref{zetatauri} (b, c)), and
the fraction of the signal within the central pixel was measured to be:
 
\begin{equation}
\Pi = \frac{\int_{0}^{r_{pixel}}\int_{0}^{2\pi}G(r,\theta)dr
 d\theta}{\int_{0}^{\infty}\int_{0}^{2\pi}G(r,\theta)dr d\theta} = 69\%.
\end{equation}

From the misspointing correction factor $\Pi$, 
it was possible to derive the LONS rate from:

\begin{equation}
p = \frac{R_{pulsed}}{R_{pulsed} + R_{nebula} + R_{LONS}},
\end{equation} 
where p is the empirical pulsed fraction, $R_{pulsed}$ is the expected rate
(table~\ref{table 2}) of Crab pulsed events, $R_{nebula}$ the expected rate
from the Crab nebula (table~\ref{table 2}) and $R_{LONS}$ is the galactic
background rate around the Crab nebula.

Finally, assuming that the contribution of the pulsed signal is negligible
compared to the background due to the nebula and the LONS we have:

\begin{equation}
p = \frac{\int
  F_{\nu}^{pulsed}\cdot \eta \cdot Q \cdot A_{ct1} \epsilon_{\gamma v} \Pi d\nu}{
  \int F_{\nu}^{nebula}\cdot \eta \cdot Q \cdot A_{ct1} \epsilon_{\gamma v}
  \Pi d\nu   + \int F_{\nu}^{LONS} \cdot \Omega \cdot \eta \cdot Q \cdot
  A_{ct1} \epsilon_{\gamma v } d\nu},
\end{equation}

where $\eta$ is the mirror reflectivity, $Q$ the quantum efficency of the PMs,
$A_{ct1}$ is the CT1 mirror area, $\Pi$ is the fraction of light
from a point source captured by the central pixel,
$\Omega$ is the solid angle subtended by the central pixel, and $\epsilon_{\gamma v}$ is a
conversion factor between number of photoelectrons and voltage measured by the
digitization card, which cancels out in the above expression. 

Therefore, using the pulsed fraction of the total Crab signal, the 
frequency integrated LONS (weighted with the frequency dependency of the 
PM quantum efficiency) is:
\begin{equation}
I_{LONS} = \int F_{\nu}^{LONS} d\nu = (4.6\pm0.1) \times 10^{12}\; ph \cdot m^{-2} \cdot 
s^{-1} \cdot sr^{-1}
\label{lons1}
\end{equation}   

To cross-check this result, we used the signal detected by CT1 from $\zeta$
Tauri in the central pixel (fig. \ref{zetatauri}): If $i_{cp}$ is the
total observed current intensity from the direction of $\zeta$ Tauri in the
central pixel, $i_{\zeta\; Tauri}$ is the expected contribution to the central
pixel current from $\zeta$ Tauri, and
$\epsilon_{\gamma i}$ is the conversion factor from photoelectrons to current
(in $\mu$A), which is calculated from

\begin{equation}
\epsilon_{\gamma i}\Pi \int
F_{\nu}^{\zeta\; Tauri}\cdot \eta \cdot Q \cdot A_{ct1} d\nu =
i_{\zeta\; Tauri} = i_{cp} - i_{LONS}\sim i_{cp}.
\end{equation}

The last approximation is valid since $\zeta$ Tauri is a very bright star
rendering the contribution from the LONS in the central pixel negligible. $F_{\nu}^{\zeta\; Tauri}$ is tabulated including atmospheric
extinction. We obtain a measured value of $\epsilon_{\gamma i} = 2.95 \times
10^{-9}~mA\cdot s\cdot phe^{-1}$. From this value we obtain the LONS by subtracting
the PSF of $\zeta$ Tauri from a square degree image centered on this star
(shown in Figure \ref{zetatauri}). The error on
the LONS is minimal near the outer wings of the PSF, since the LONS is
dominant in that area. An annular region is used to estimate the LONS, giving

\begin{equation}
I_{LONS} = \int F_{\nu}^{LONS} d\nu \sim (4.3\pm0.1) \times 10^{12}  ph \cdot m^{-2} \cdot s^{-1}\cdot sr^{-1},
\label{lons2}
\end{equation}
after normalising to the solid angle subtended by the selected annular region.

The two values for the galactic $I_{LONS}$ (eq. \ref{lons1} $\&$ \ref{lons2})
calculated by two independent methods agree within $\sim 7$\% and are
$\sim$~2.7 times higher than the previous measurement \cite{Razmik},
used for our theoretical predictions as we expected. The LONS intensity
calculated from the pulsed fraction is the most accurate, being the correction
factor $\Pi$ uncertainty the highest, since the LONS determination from $\zeta$
Tauri also includes the uncertainty in the background region around the star.

\section{Sensitivity for Pulsed Detection in Optical}

\begin{figure}[h]
\epsfysize=12cm 
\epsfxsize=12cm 



\hspace{0.5cm}
\epsfbox{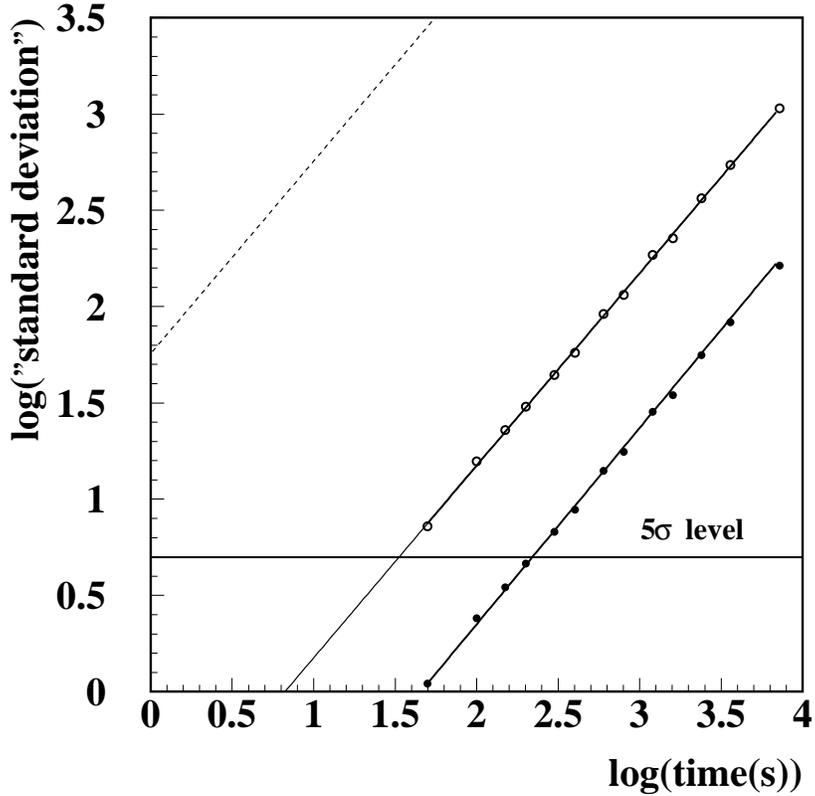}



\caption[h]{\it log of "standard deviation" {\bf s} of the signal related to the background
  versus observation time. The CT1 empirical data are represented
  with filled bullets while the empty ones represent the extrapolation for the
  MAGIC telescope. The horizontal line shows the "5$\sigma$" level.  }
\label{sensitivity}
\end{figure}

The sensitivity of CT1 telescope to detect the Crab optical pulse can be
expressed by the "standard deviation" {\bf s} of the signal relative to the
behavior of the reduced $\chi^2$ at frequencies different from the Crab frequency:

\begin{equation}
s = \frac{\chi^{2}_{red} - \mu_{background}}{\sigma_{background}}
\end{equation}
for each observation time interval, where $\mu_{background}$ is the mean value
of $\chi^{2}_{red}$ at such offset frequencies and $\sigma_{background}$ the
corresponding standard
deviation, which is constant for all selected times $T$. 
The total 2 hours observation time was split into $L=7200\;s/T$ independent
datasets, each of length $T$, from which the average value of the statistic
$s$ was calculated for $T=50$ s to 7200~s in steps of 50~s. It can be shown
that this average value of $s$ is:   

\begin{equation}
s \sim p^2 \cdot T \cdot g(k,f_{s}), 
\end{equation}
where $k$ is the number of bins in the phaseogram, 
$f_{s}$ the signal pulse shape and g(k,$f_{s}$)
a function depending on these quantities.
Note that $s$ is not a Gaussian standard deviation, since the number $k$ is too
small, but we expect that $<s> \propto \sigma^{2}$, where $\sigma \sim
p\cdot\sqrt{T}$ would be the Gaussian significance of a DC detection of the
same signal.
Figure \ref{sensitivity} shows a graph for $log<s>$ vs. logT, showing
a linear relationship between $s$ and $T$ as predicted.
 
Using the calculated value of the galactic LONS from the pulse analysis, we
estimate the required time for MAGIC to detect the Crab pulsar in the optical,
given the much larger mirror area, improved quantum efficiency, and pixel
size (0.1$^o$).
By scaling the graph of log$<s>$ vs. logT accordingly, Figure
\ref{sensitivity} shows that MAGIC can detect the Crab pulsar in 30~s at
the 5$\sigma$ level. The line on top shows the prediction for a PM without inherent fluctuations (eq. 3), using the galactic LONS calculated.

\section{Conclusions}

The Crab pulsar was measured in the optical with the HEGRA CT1
telescope (the smallest \^{C}erenkov telescope making this detection). 
A 5$\sigma$ detection requires an observation time of 200~seconds. 

Since the design of the CT1 PM base is very similar to that of MAGIC, we
extrapolate CT1 results and get a 5$\sigma$ detection with MAGIC ($\sim$
25 times larger dish area than CT1) in a substantially shorter time of
30~sec. Considering that it is possible to reduce the LONS using an array of
HPD's, combined with the superior PSF of MAGIC, it could still be possible to
resolve single pulses.

The use of a PM base with the technical modifications described above is incompatible with normal VHE observations. The modification could
only be considered for the central pixel of the telescope camera. 

The galactic LONS background on the surroundings of the Crab region was
calculated using the Crab pulsed fraction. 
Only $7$\% disagreement was found with a
second method using a bright reference star and a different background region
(still within the camera field-of-view). Our results are however on average 2.7 times
higher than the extragalactic LONS from \cite{Razmik}.
The advantage of this new method is that it relies on the stability of
the clock, and is independent of other sky-techniques in the DC mode,
provided that the PSF is known.
The determination of the LONS for the Crab region is important, since the
threshold energy determination and the spectrum depend on the level of the
LONS.   

Simultaneous Crab pulsar optical data and \^{C}erenkov VHE observations could
be very important for monitoring atmospheric extinction.
In fact, high quality observations should show a very specific zenith angle
dependency of the pulsed fraction. Thus, the "seeing condition" at a given
time may also be 
derived from 
the observed pulsed fraction corrected
for atmospheric extinction. Furthermore, apart from using optical observations
to provide a timing ephemeris for VHE observations.

We can also monitor deviations from the normal cubic law for pulsar spindown
(assuming the pulsar braking index is detectable) as a result of timing noise and glitches.
Thus, frequent monitoring of the Crab by MAGIC (for calibration
purposes) should also improve the chance of detecting glitches. In fact,
very specific software can be written to test glitching behavior
during observations. A specially designed central pixel detector with high
quantum efficiency, and relatively low noise will be very useful for
monitoring the state of optical emission from $\gamma$-ray
sources such as cataclysmic variables, X-ray binaries, microquasars, AGNs and
possibly GRBs.

{\bf Acknowledgments.}
 
 We want to thank David Smith for his help with the technical
  details of the optical setup. We are also grateful to Razmik Mirzoyan,
  Eckart Lorenz, Manel Martinez, H.G. Boerst and Georges Blanchot for
  essential technical help. Thanks to Jose Luis Contreras, Juan Abel Barrio
  and Mathew Holleran for lots of discussions and corrections.













\begin{thebibliography}{00}






\bibitem{MAGIC} 
Barrio J.A., et al. 1998, ``The Magic Telescope'', design study, MPI-PhE/98-5 (1998) 
%
\bibitem{carram}
Carrami\~nana A., Cadez A. and Zwitter T. ApJ, 542, (2000) 974-977 
%
\bibitem{jager}
De Jager O.C., Konopelko A., Raubenheimer B.C and Visser B., 27$^th$ ICRC
proceedings, Hamburg 26-30 June (2001)
%
\bibitem{celeste_crab}
De Naurois M. et al., CELESTE Coll. (P. Espigat, F. Münz, A. Volte). ApJ 566 (2002) 343-357
%
\bibitem{tdas02} 
De O\~na-Wilhelmi E., J. Cortina and O.C. de Jager. MAGIC-TDAS 02-09 (2002)
%
\bibitem{tempo}
Manchester R.N. et al. MNRAS, 328 (2001) 17-35
%
\bibitem{Razmik}
Mirzoyan R. and Lorenz E. Internal report of the HEGRA
collaboration. MPI-PhE/94-35 (1994)
%

\bibitem{CT1}
Mirzoyan R. et al, Nucl. Instr. and Meth. A, 351 (1994) 513.
%
\bibitem{whipple_crab} 
Srinivisan R. et al., in Proc. ``Towards a Major Atmospheric 
\^{C}erenkov Detector V'', p. 51, ed. O.C. de Jager, 1997.
%

\end{thebibliography}
\end{document}